\documentclass{article}
\usepackage{spconf,amsmath,graphicx}

\title{Phone and speaker spatial organization \\in self-supervised speech representations}
\name{Pablo Riera $^{\star\dagger}\qquad$ Manuela Cerdeiro $^{\star\dagger}\qquad$ Leonardo Pepino $^{\star\dagger}\qquad$ Luciana Ferrer $^{\star}$ \vspace{-0.3cm} \thanks{This work was supported by a Google Faculty Research Award, 2019.
Correspondence: priera@dc.uba.ar}}
\address{
$^{\star}$Instituto de Investigaci\'on en Ciencias de la Computaci\'on (ICC), CONICET-UBA, Argentina \\
$^{\dagger}$Departamento de Computaci\'on, FCEyN, Universidad de Buenos Aires (UBA), Argentina}
\begin{document}
\ninept
\maketitle
\begin{abstract}
Self-supervised representations of speech are currently being widely used for a large number of applications. Recently, some efforts have been made in trying to analyze the type of information present in each of these representations. Most such work uses downstream models to test whether the representations can be successfully used for a specific task. The downstream models, though, typically perform nonlinear operations on the representation extracting information that may not have been readily available in the original representation. In this work, we analyze the spatial organization of phone and speaker information in several state-of-the-art speech representations using methods that do not require a downstream model. We measure how different layers encode basic acoustic parameters such as formants and pitch using representation similarity analysis. Further, we study the extent to which each representation clusters the speech samples by phone or speaker classes using non-parametric statistical testing. Our results indicate that models represent these speech attributes differently depending on the target task used during pretraining.

%Analyzing representations generated by the layers of deep neural networks helps make their black-box behavior more interpretable. Probing the network's hidden layers on different tasks can provide insights about the information encoded in each layer, with the side effect that the downstream model might distort the structure of the representation, disentangling some of the initially encoded information (like speaker identity). This work investigates preserving information in the internal layers of self-supervised speech representations without using a prediction task. We use representation similarity analysis and statistical techniques to measure how the representations organize speech signals. We measure how layers encode basic acoustic parameters such as formants and pitch, and we also test if the representations tend to cluster the sounds based on phone or speaker classes. The overall results indicate that models represent speech attributes differently depending on the target used during pretraining.

\end{abstract}
\begin{keywords}
Speech Representations, Self-Supervised Learning, Representation Analyses
\end{keywords}
\section{Introduction}
\label{sec:intro}

In recent years, many new deep-learning-based speech representations have been proposed and used for a variety of applications. Most of these models are trained with self-supervised approaches~\cite{liu2022audio}, making it hard to understand which information is being preserved in the resulting representations. Consequently, several recent works have focused on analyzing how the properties of a speech signal are encoded in these representations \cite{ma2021probing,de2022probing}. These works rely on different techniques for probing speech representations. Some of them are based on evaluating downstream tasks involving training a machine learning model that takes the representations from the pre-trained model as input. The output labels are characteristics derived from the speech signal, such as phone class \cite{ma2021probing}, pronunciation quality, fluency, or other speech properties \cite{shah2021all}.
These approaches tell us whether a certain representation encodes a particular speech property, but they do not specifically tackle the question of the structure of the representation in the embedding space. %Analyzing the structure means understanding how and why the speech signals are mapped or distributed in the embedding space. 

%For example, an interesting situation of organization of speech properties arises in the case of phones and speaker classes. Given a representation, it is not possible to cluster phones and speaker classes at the same time as the set of phones are shared by all the speakers with the same language. Nonetheless, by scaling or discarding spatial dimensions, it is possible to favor either phones or speakers. This may result in a good performance when using a downstream model in the tasks of phone recognition or speaker identification but does not answer the question of the spatial organization of the classes in the representation. In this work, we avoid using downstream models and focus on different ways of representation analysis.

For example, an interesting situation arises when analyzing phone and speaker classes. A given representation cannot cluster well both phones and speakers since clustering by speaker implies that phone information must be ignored and conversely. Nonetheless, by non-linearly transforming the embeddings with a specific downstream model for each case, it may be possible to classify both phones and speakers with the same embeddings. In this work, though, our goal is to understand the underlying organization of the embeddings as they come out of the model. For this reason, we avoid the use of downstream models.

% Dimensionality reduction techniques allow us to visualize multidimensional representation in a two-dimensional space and inspect how different data properties distribute or cluster together. These techniques are beneficial for exploratory analysis, but they depend on tuning hyper-parameters which could generate biased conclusions.  

One way of analyzing the structure of a representation is by comparing it with another one. A comparison of two representations can be made using methods from Representation Similarity Analysis (RSA) \cite{kornblith2019similarity}. One of these methods is centered kernel alignment (CKA), which measures the similarity of the geometric structure of two representations and it has been used to identify correspondences in representations that were trained using different initializations. RSA techniques have been used to compare representations extracted from different hidden layers within the same neural network. In \cite{pasad2021layer}, they found that wav2vec2.0 layers best encoded phonetic information in the middle layers. Using CKA, \cite{chung2021similarity} found that the learning objective of self-supervised speech models affects the similarity more than the architecture does. %We would like to highlight that finding that two speech representations are dissimilar does not imply one will perform better than the other in a downstream task. They could be representing the target attribute in different ways or at different levels of abstraction. In fact, combining representations could lead to improved performance over using the single best representation for a certain task \cite{wu2022ability}. The disadvantage of RSA methods is that they do not explain why any two representations are dissimilar.

A different approach for analyzing a representation structure involves measuring how members of the same class (for example, phones) are clustered in that space. One way of computing a metric for this is by using an ABX discrimination task \cite{schatz2013evaluating} where every sample X is classified as being of class A or not by comparing the distances in the representation space of samples A, B, with X, with B in a different class from A. This method has been used to measure the intrinsic quality of a speech representation to perform a certain classification task \cite{algayres2020evaluating,carlin2011rapid}. %The method allows rapid evaluations of these representations \cite{carlin2011rapid}.

In this work, we analyze how different self-supervised speech models represent phone and speaker information using methods that do not require downstream models. First, we analyze the similarity of a set of acoustic features with the representations generated by the models' layers using linear CKA. Second, using non-parametric statistical testing, we measure how speech samples from the same class (phone or speaker) cluster together in the space. Specifically, given a speech sample, we measure to what extent its nearest neighbors belong to the same class using a multivariate Wilcoxon-Mann-Whitney (WMW) test. Our analysis suggests that representations with the same learning strategy (i.e., trained for the same self-supervised task) tend to represent the same type of information, regardless of their specific architecture.

% Our analysis is carried on using different representations, both classic features like melspectrogram and MFCC, and also modern self-supervised models, like wav2vec2.0 and others described below. The models based on neural networks generally have many hidden layers that can be analysed separately and it has been shown that speech properties are encoded differently in each layer \cite{shah2021all}\cite{pasad2021layer}. 

% In this work we perform two analysis to investigate which speech properties are encoded in different layers of speech representation models without fitting downstream models.

\section{Speech Representations}
\label{sec:reps}

\begin{table*}[t]
\centering
\begin{tabular}{ccccccc}
Model        & Dataset & Input Format        & Encoder              & Loss              &  Target \\ \hline
Mockingjay \cite{liu2020mockingjay}  & LS 360 hr & mel-spectrogram     & Transformer          & L1 loss           &  mel-spectrogram       \\
DeCoAR2 \cite{ling2020decoar}      & LS 960 hr & log mel-spectrogram & Transformer          & L1 loss           & mel-spectrogram \\
HuBERT Base \cite{hsu2021hubert}   & LS 960 hr & raw waveform        & 1D CNN + Transformer & Contrastive loss & K-Means MFCC \\
WavLM Base+ \cite{chen2022wavlm} & Mix 94k hr & raw waveform        & 1D CNN + Transformer & Contrastive loss    & K-Means MFCC   \\
wav2vec 2.0 Base \cite{baevski2020wav2vec} & LS 960 hr & raw waveform        & 1D CNN + Transformer & InfoNCE &  Internal \\
data2vec \cite{baevski2022data2vec}     & LS 960 hr & raw waveform        & 1D CNN + Transformer & L1/L2 Loss        & Internal \\
\hline

\end{tabular}
\caption{Self-supervised speech models considered in the analysis and their main characteristics. Here LS means LibriSeech dataset, and Mix is a dataset introduced in the WavLM work that combines different datasets.}
\label{tab:tab1}
\end{table*}

The success of transformer models and self-supervised learning in NLP inspired many of the recently proposed neural-network-based speech representations. Specifically, most of these models are based on the masked language modeling pretext task from BERT \cite{devlin2018bert}, which consists in reconstructing masked regions from the input signal.
In this work, we analyze a variety of self-supervised speech representations. Table \ref{tab:tab1} shows the list of models used in this work, including their distinguishing characteristics.

Mockingjay \cite{liu2020mockingjay} follows the same transformer architecture used in BERT (12 layers of 768 dimensions and 12 attention heads). The model masks a portion of the mel-spectrogram input features and then learns to predict the masked frames using the transformer's last layer and L1 loss. %, samples 15\% of the input frames and applies masks of length 3 to each of them. 
%The main difference with BERT is that the inputs and targets are continuous speech instead of tokens representing words. 
%This setup avoids the quantization of the input speech and replaces the cross-entropy loss with an L1 loss. 
Similarly, DeCoAR2 \cite{ling2020decoar} learns to reconstruct a masked gap from the input log mel-spectrogram using a transformer encoder, but they include a Gumbel-Softmax quantization layer before the prediction head. The authors show that the quantization layer boosts performance when the model is fine-tuned for ASR. 

Using the same masked language modeling pretext task, in HuBERT \cite{hsu2021hubert}, the targets are generated by k-means clustering of MFCCs. WavLM \cite{chen2022wavlm} follows the same strategy, but they introduce a speech-denoising task by generating mixtures of speech and noise. The pretext task predicts the targets generated using the unmasked clean speech from the noisy masked speech.

In wav2vec 2.0 \cite{baevski2020wav2vec}, a CNN encoder generates the inputs to the transformer applied to the speech waveform, and they use Gumbel-Softmax quantization \cite{jang2016categorical} to generate the targets and minimize a contrastive loss.
%The masked language modeling task is used as a self-supervised pretext task, and the InfoNCE \cite{oord2018representation} contrastive loss is applied. This loss encourages the model to output representations of the masked regions that are similar to the original quantized ones but also dissimilar to other regions of the input speech. 
In data2vec \cite{baevski2022data2vec}, the model itself generates the contextualized targets. To train data2vec, the unmasked speech signal is forwarded to a teacher model, which has an exponential moving average of the weights of the student model, and the student model has to predict the activations of the teacher model from the masked speech signal by minimizing a regression loss.

Finally, we also include, for the analysis, classic speech features such as 13-dimensional MFCC with their deltas and double deltas, 80 bins mel-spectrogram, and Kaldi's filter bank features \cite{povey2011kaldi}.

All of the transformer-based architectures mentioned above have 12 layers. In the results presented below, we also include the projection layer that adapts the input to the transformer block as layer zero. 
In this paper, we consider every instance of a phone as a sample and the neural representations and acoustic features are averaged over all the frames in each phone for our analyses. %The representations were computed using the open-source repository available in https://github.com/s3prl/s3prl \cite{s3prl}. 

\begin{figure}[htb]
\begin{minipage}[b]{1.0\linewidth}
  \centering
  \centerline{\includegraphics[width=8.5cm]{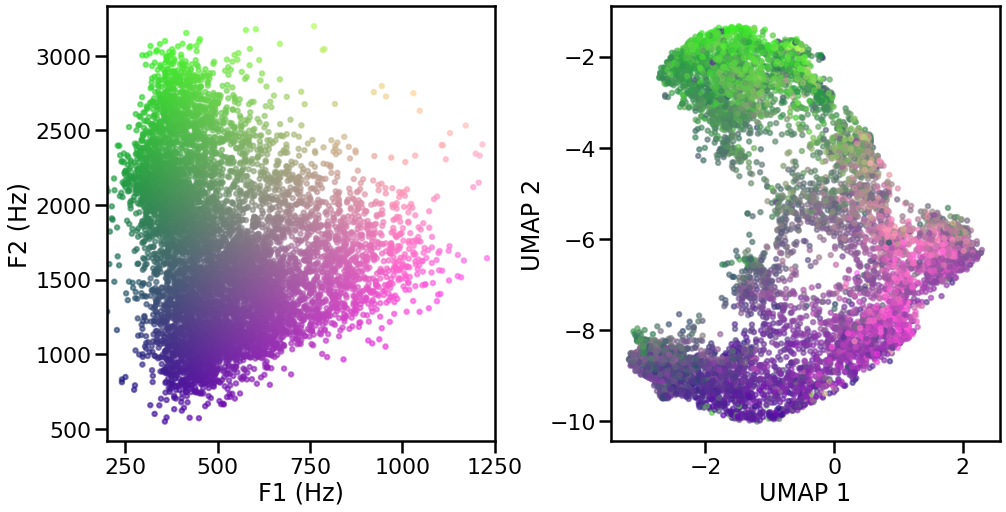}}
\vspace{-5pt}
\caption{Left: F1 vs F2 values for vowels from the L2Arctic dataset. %organized by color in the F1 and F2 space. 
Right: UMAP visualization of the 4th layer of WavLM for the same vowels. The color of each point is determined based on the F1-F2 values.}
\label{fig:umap1}
\end{minipage}
\end{figure}

\begin{figure}[htb]
\begin{minipage}[b]{1.0\linewidth}
  \centering
  \centerline{\includegraphics[width=9cm]{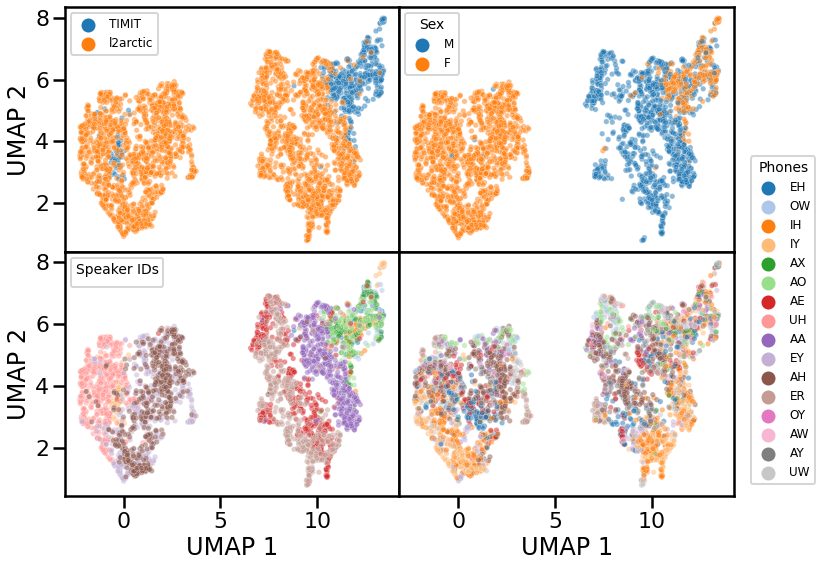}}
\vspace{-6pt}
\caption{UMAP visualization of the vowels representation using the 12th layer of DeCoAR2 for a subset of 12 speakers, coloring the samples by the labels indicated in each figure. %This representation maintains a speaker organization rather than a phone one. 
%Top Left: colored by the database. Top Right: colored by sex. Bottom Left: colored by speaker id. Bottom Right: colored by vowel.
}
\label{fig:umap2}
\end{minipage}

\end{figure}

\section{Methods and metrics}
\label{sec:mam}
In this work, we use two methods to analyze the spatial structure of the representation. 
The first method measures the similarity between the representation under analysis and acoustic representations like formants. In Figure \ref{fig:umap1}, we find a visualization that motivates this analysis. In the left plot, we see the average F1 and F2 values for each vowel in the L2Arctic dataset. The colors are determined based on the coordinates. In the right plot, we see the UMAP %Uniform Manifold Approximation and Projection)
\cite{McInnes2018umap} 2D visualization of the 4th layer of the WavLM model (averaged over all frames within each vowel) for the same samples using the same color for each point as in the left plot. 
%It is possible to see the points preserving the color organization and showing some space continuity but with some differences. 
We can see that, while the overall structure differs, neighbors appear to be preserved across the two spaces.

To capture this phenomenon, we measure the similarity between the acoustic and neural representations using Linear Central Kernel Alignment (CKA) from the literature on representation similarity analysis \cite{kornblith2019similarity}. Given a matrix $X$ where the rows correspond to the samples (phones in our case) and the columns to the representation dimensions, this index computes the correlation of the dot-product self-similarity matrices for each representation $X$ and $Y$: $CKA(X,Y) = \mathrm{corr}( \mathrm{vec}(XX^T),\mathrm{vec}(YY^T))$, where $\mathrm{vec}$ indicates the flattening of the matrix. This metric allows us to compare representations with different dimensionality. Linear CKA takes values between 0 and 1 and measures to which degree the pairwise distance between samples is preserved between both representations. %In figure \ref{fig:umap1}, we see the preservation of local structures, but a global distortion penalizes the CKA index, which in this case takes a value of near 0.4.
%In this work, the samples are given by phones and the vectors used to compute CKA are the averages of the selected representation over all frames within the phone.
The CKA index between the two representations in Figure \ref{fig:umap1} is 0.5. While the local structure appears to be mostly preserved, the change in global structure reduces the value of CKA.

The second method used for analysis in this paper quantifies how speech segments are clustered depending on their attributes, like phone or speaker class. This is illustrated in figure \ref{fig:umap2}, which shows the UMAP projections of the 12th layer of the DeCoAR2 model. In each plot, each point corresponds to a vowel, and its color corresponds to a different characteristic of the sample: the database, the gender, the speaker, and the phone. We can see that the samples from the same speaker, dataset, and gender, for this representation, tend to cluster together. On the other hand, samples from the same phone but different speakers do not cluster together, though phones from the same speaker tend to appear nearby. For our analysis, we would like to measure whether the points belonging to a class are organized in clusters or distributed in different regions. To this end, we use a multivariate version of the Wilcoxon-Mann-Whitney (WMW) test. For each point $x$ of a given class, we create the distance rankings to all other points and compute the statistic:
\begin{equation}
    U_x= \frac{\max(R_{1}-{n_{1}(n_{1}+1) \over 2}, R_{2}-{n_{2}(n_{2}+1) \over 2})}{n_1 n_2},
\end{equation}
where $R_1$ is the sum of the rankings for the $n_1$ points of the same class as $x$, and $R_2$ is the sum of the rankings for the $n_2$ points of a different class than $x$. This statistic increases when the points of the same class as $x$ are closer to $x$ than the points of all other classes. Finally, we compute the mean of $U_x$ for all the points $x$, which we will call $AvgU$. This statistic lies in the range $0.5 - 1$, and the univariate WMW is equivalent to the AUC \cite{yan2003optimizing}. The computation of the presented statistic is a variant of the multivariate statistic presented in \cite{liu2022generalized}.

% \section{Databases}

The datasets used in our analyses are the L2Arctic \cite{zhao2018l2arctic}, and TIMIT \cite{timit} datasets, both of which have manually-checked phone alignments, and use the ARPABET phone dictionary. A significant difference between these datasets is that L2Arctic is composed of English utterances from non-native speakers, while TIMIT only has native English speakers. % Further, L2Arctic has fewer speakers but more utterances per speaker.

\section{Acoustic vs Neural representations}

Our first analysis compares basic acoustic features, F0, F1, F2, and spectral centroid \cite{martin2021speech}, with the various neural speech representations in Table \ref{tab:tab1}. The goal is to analyze whether neural representations are organized similarly for vowels as these standard acoustic features. We separate the analysis of these variables into two groups to test their representation correspondence separately. We group F0 and spectral centroid, which are features known to contain information about the speaker identity, and F1 and F2, which correlate with the vowel identity (see the left plot in Figure \ref{fig:umap1}). 

\begin{figure}[htb]
\begin{minipage}[b]{1.0\linewidth}
  \centering
  \centerline{\includegraphics[width=8.5cm]{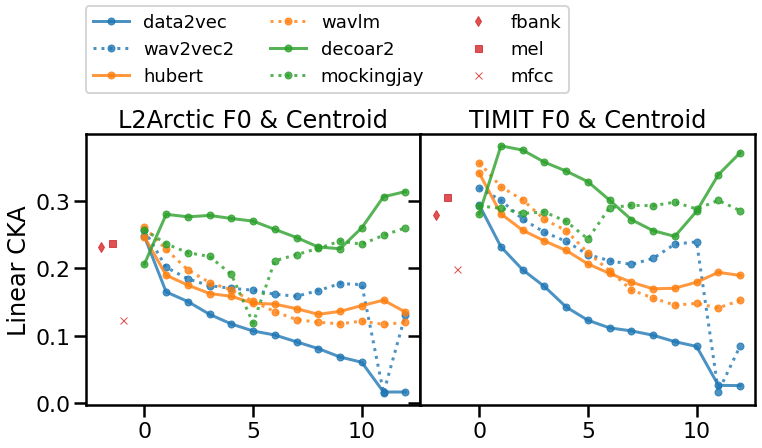}}
    \centerline{\includegraphics[width=8.5cm]{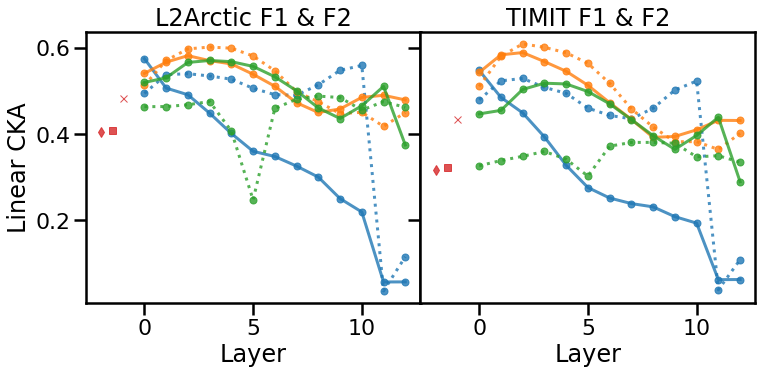}}
\vspace{-6pt}
\caption{Linear Central Kernel Alignment values of acoustic features (F0 \& centroid, or F1 \& F2) vs. audio representations for L2Arctic and TIMIT vowels. 
%Top: Acoustic features F0 and centroid. Bottom: F1 and F2. 
Curves with the same color indicate neural models used similar pretraining targets. The non-neural features (red markers) are positioned on the left in separate locations for easier visualization.}
\label{fig:result1}
\end{minipage}
\end{figure}

Figure \ref{fig:result1} shows the results for L2Arctic and TIMIT vowels. We measure Linear CKA between F0 and centroid features (top) and F1 and F2 features (bottom) versus six neural representations with 1 projection layer (layer 0 in our plots) and 12 transformer layers, and 3 classic speech features: MFCC, mel-spectrogram, and Kaldi's filter bank. The trends for L2Arctic and TIMIT are similar, though L2Arctic shows smaller CKA values for the F0 and centroid case. This could be happening due to inaccurate F0 values for some of the L2Arctic data. Alternatively, this may suggest that the pre-trained models do not represent these acoustic features as well for native (TIMIT) than for non-native speakers (L2Arctic). 

For the F0 and centroid features, we can see that the WavLM, wav2vec2.0, HuBERT, and data2vec representations preserve the acoustic information structure in the initial layers and partially lose it  throughout the successive layers. The trend is different for DeCoAR2, and Mockingjay, where the last layers have similar or greater CKA values than the first ones. These models' targets are the same as their input features. Interestingly HuBERT and WavLM behave similarly, which could be related to the fact that both models use the same target type: quantized MFCCs. The last two layers of wav2vec2.0 and data2vec also have similar behavior. This could be due to the fact that these models' reconstruction targets are internally learned, and the last layers may not necessarily be organized by acoustics features. In the case of Mockingjay, a dip can be seen in the 5th layer, which may be related to the architecture having different dimension size in the linear projection matrix for that layer. Overall, there appears to be a strong relationship between the organization of the last layer and the model's learning objective during pretraining.

For F1 and F2, we can see that the similarity value is larger than for F0 and centroid. This behavior is expected since these representations are good at encoding phone information, in this case, vowels. While some representations like data2vec and Mockingjay present the same curve trend as in the F0 and centroid case, HuBERT, WavLM, wav2vec2.0, and DeCoAR2 have peaks of similarity between the layers 2 to 4. These layers appear to preserve the phone information best, as shown in the next section's results. We also analyzed the use of articulation features and obtained similar results to those of F1 and F2 (results not shown due to lack of space).

\begin{figure}[htb]
\begin{minipage}[b]{1.0\linewidth}
  \centering
  \centerline{\includegraphics[width=8.8cm]{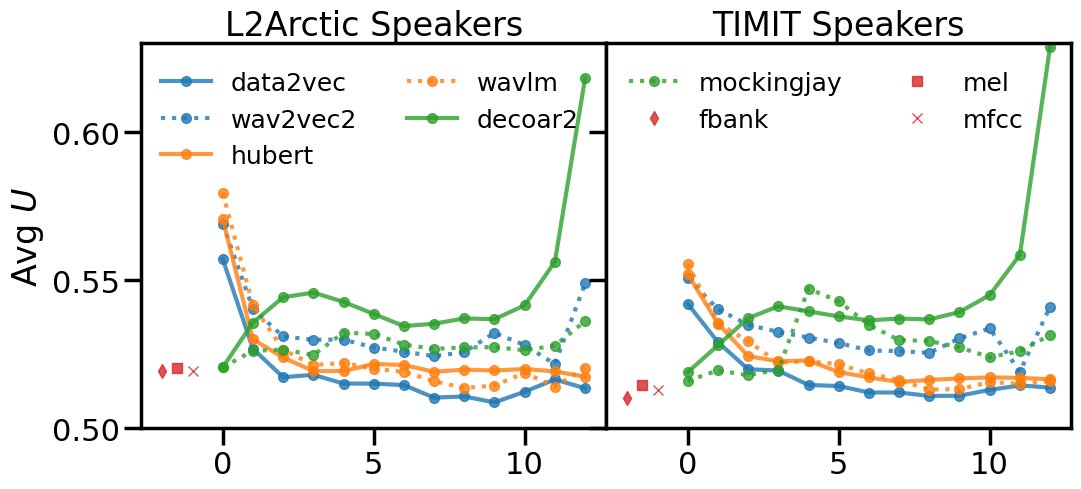}}
    \centerline{\includegraphics[width=8.8cm]{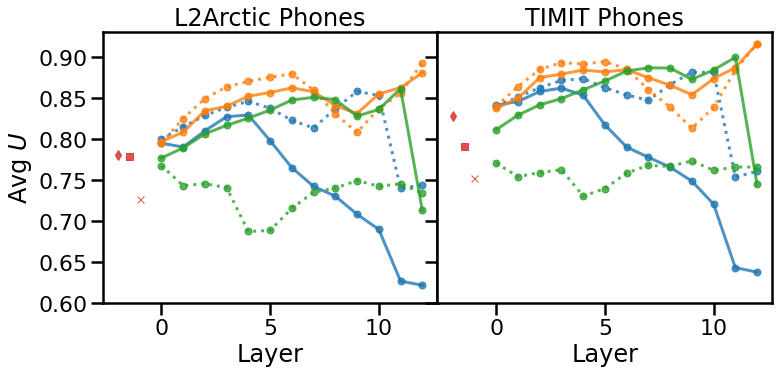}}
\vspace{-5pt}
\caption{Multivariate Wilcoxon-Mann-Whitney (WMW) statistic (Avg U) for speakers (top) and phones (bottom) cases in L2Arctic (left) and TIMIT (right). For each representation layer, the score indicates how well the points of a given class are separated from the other classes. Curves with the same color indicate neural models that use similar pretraining targets. The non-neural features (red markers) are positioned on the left.}
\label{fig:result2}
\end{minipage}
\end{figure}

\begin{figure}[t]
\begin{minipage}[b]{1.0\linewidth}
  \centering
  \centerline{\includegraphics[width=8.5cm]{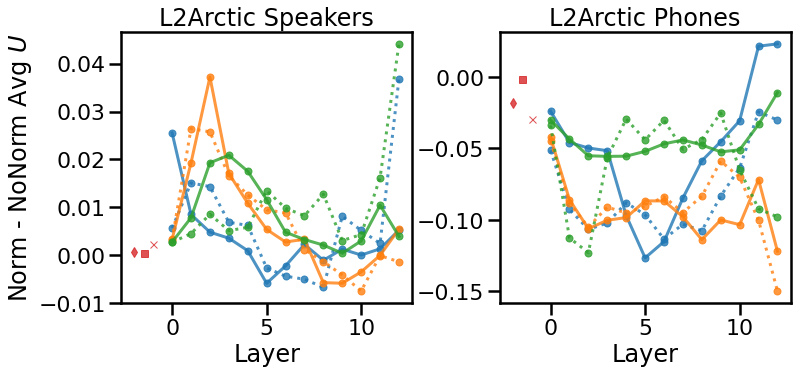}}
\vspace{-6pt}
\caption{Comparison of results for mean and variance normalization. Difference of $AvgU$ for speaker and phone cases in L2Arctic.}
\label{fig:result3}
\end{minipage}
\end{figure}

\section{Encoding of speaker and phone classes}
% \vspace{-1.0em}

For the second analysis, we want to measure if a point of a given class (speaker or phone) is near other points of the same class, for which we use the $AvgU$ metric defined in section \ref{sec:mam}. Figure \ref{fig:result2} shows the $AvgU$ scores for speaker (top) and phone (bottom) classes for each representation layer for L2Arctic (left) and TIMIT (right). Both databases present similar results, with TIMIT values being somewhat higher in the case of phones. This shows some stability of the representations to non-native pronunciations, supporting the hypothesis that the differences between L2Arctic and TIMIT in Figure \ref{fig:result1} were due to inaccuracies in the features rather than a degradation in the representations for non-native speakers. 
% We compute the average multivariate Wilcoxon-Mann-Whitney (WMW) statistic $U$. %The total number of samples had to be restricted for this analysis because it involves computing distances between all pairs of points. Hence, we limit the number of utterances in L2Arctic and speakers in TIMIT to reach roughly 20000 phone samples in each database.

For the metric computed by speaker (top plots), we can see that data2vec, WavLM, and HuBERT, have larger values in their first layers and then decline, while DeCoAR2 and wav2vec2.0 increase the value in their last layers. We see that the phone $AvgU$ is larger than the speaker $AvgU$ for any specific representation, meaning the points tend to be clustered into phone classes. This is expected as the learned task for these models is to predict a masked target using the context, which can be improved by learning the phone regularities in the language. Interestingly, we see a clear trade-off between speaker and phone $AvgU$ values for the same representation. If a representation clusters speakers, the phones will be dispersed between the clusters corresponding to each speaker. If, on the other hand, a certain representation clusters phones together, then the points corresponding to different speakers will overlap. Figure \ref{fig:umap2} shows a case where the speakers are clustered, and the phones are segregated. 

For the metric computed by phone (bottom plots in Figure \label{fig:result2}), we observe that the peak values depend on the model. For models like WavLM and HuBERT there is a peak in the last layers. This may be explained by the fact that these models are trained to predict quantized MFCCs, which are known to be good for phone classification. We see that the final layers have lower values for DeCoAR2, Mockingjay, wav2vec2.0, and data2vec. These models are trained to predict either mel-spectrogram or internal representations where the phone information may not be as readily available as in MFCCs. We can also see that the models that use internally-generated targets, wav2vec2.0 and data2vec, have distinct last layer values. In particular, data2vec value is the lowest, meaning that the targets used to train it are not well organized by phone.  %by it are less representative of the task.

Another interesting result occurs when the representations are normalized. The normalization is done by taking all the phones in a database, subtracting the mean from each representation dimension, and dividing by the standard deviation. Figure \ref{fig:result3} shows the difference in $AvgU$ value for normalized versus unnormalized representations. A slightly positive value for the speakers indicates that the normalization favored speaker clusterization. For the phone case, negative values indicate the opposite.

%To compare our results with downstream task performances, we compute the correlation between the maximum value of the $AvgU$ and the performances reported in the superb benchmark \cite{yang2021superb}. For the $AvgU$ of phones, we found that several superb tasks have significant correlations, % many significant correlations with several tasks in the benchmark,
%we report the value for the phoneme recognition task, which gives a Pearson correlation of 0.84 ($p=0.018$). For the case of speaker $AvgU$, we found only one high correlation, which corresponds to the speaker identification task, with a Spearman correlation of 0.75 ($p=0.05$).

Finally, we can compare the values of $AvgU$ with the performance obtained on downstream tasks. To this end, we use the performances reported in the SUPERB benchmark \cite{yang2021superb}\cite{superb} computing their correlation with the maximum $AvgU$ over the layers for each downstream task, over the self-supervised models analyzed in this paper. For the $AvgU$ of phones, we obtained a correlation of 0.84 ($p=0.018$) with the performance on the phone recognition task. Similarly, for the $AvgU$ of speakers, the correlation with the performance on the speaker identification task was 0.75 ($p=0.05$). This indicates that each $AvgU$ is able to predict which of the self-supervised models is best for the class for which they are computed.

%Interestingly, the classic features MFCC, mel-spectrogram, and the filter bank in general have similar scores to those of the first layers for some models in the phones case.

%We also tested using a cosine metric instead of euclidean in the WMW test, but the curves' trends remained similar. This can be explained as the vector points in the representation tend to have a similar length, which makes the euclidean metric correlate with the cosine metric.

\section{Conclusions}

In this work, we analyze phone and speaker spatial organization in speech representations obtained from self-supervised models using two techniques that do not rely on training downstream models. In the first case, we compare acoustic features and neural representations using the representational similarity index, linear CKA. Next, we introduce a multivariate test to measure whether phones and speakers are clustered in the representations. We observe that the performance on these tasks depends heavily on what type of learning objective was used in the pretraining. Our analyses suggest that the first few layers encode the information available in the transformer input while the middle layers organize the space in a way that helps phone representation, and the last layers favor a representation that matches the target task used for pretraining.

%representations and their layers preserve more or less the acoustic organization of F0 and centroid, or F1 and F2. We also see that classical features like MFCC, mel-spectrogram and filter bank are less capable of maintaining the acoustic spaces.

%\vfill
%\pagebreak
\vfill\pagebreak

\bibliographystyle{IEEEtran}
\bibliography{refs}

\end{document}